\documentclass{revtex4-1}
\usepackage[latin9]{inputenc}
\usepackage{verbatim}
\usepackage{amsmath}
\usepackage{amssymb}
\usepackage{graphicx}
\begin{document}

\title{Spontaneous Super-Rotation on Planets}

\author{Masahiro Morikawa\thanks{{}hiro@phys.ocha.ac.jp} }

\affiliation{Department of Physics, Ochanomizu University, 2-1-1 Otsuka, Bunkyo,
Tokyo 112-8610, Japan}

\date{\today}
\begin{abstract}
super-rotations of the planetary atmosphere are reconsidered from
the dynamical point of view. In particular, we emphasize that the
super-rotation appears spontaneously without any explicit force. Although
the super-rotation violates the bilateral symmetry (east-west reflection
symmetry) of the system, this violation is spontaneous. Constructing
a minimal model that derives the super-rotation, we clarify the condition
for the super-rotation to appear. We find that the flow is always
determined autonomously so that the flow speed becomes maximum or
the temperature difference smallest. After constraining the parameters
of the model from observations, we compare our model with the others
most of which demand the explicit symmetry violation due to the planetary
rotation. 

\end{abstract}

\pacs{33.15.Ta}

\keywords{Suggested keywords}

\maketitle

\tableofcontents

\section{Introduction\label{sec:Introduction} }

We reconsider the super-rotation flow pattern typically observed in
the Venus atmosphere. It is really curious that the flow speed far
exceeds the rotation speed of the Venus. There have been much research
before including elaborate numerical calculations and various proposals
to yield strong Westward zonal flow. However, the decisive mechanism
of the super-rotation has not yet been clarified\cite{Bengtsson2013}.
Basic problems include (1) why the flow velocity far exceeds the planetary
rotation speed is possible, (2) what maintains this super-rotation,
and (3) how the flow violates the natural symmetry along the axes
connecting the day point (nearest to the Sun) and the night point
(farthest to the Sun) of the planet. 

The super-rotation (SR) is generally an entirely coherent flow on
the planet in one rotational direction often faster than or comparable
to the planetary rotation speed. However, if we also include the flow
locally coherent in one rotational direction as SR, we observe many
examples; planetary atmospheres of Venus and Titan, Earth and Jupiter
jet streams, Solar differential rotation,... The above SR problem
should be considered in this wide point of view emphasizing the generality
of SR. 

\begin{figure}
\includegraphics{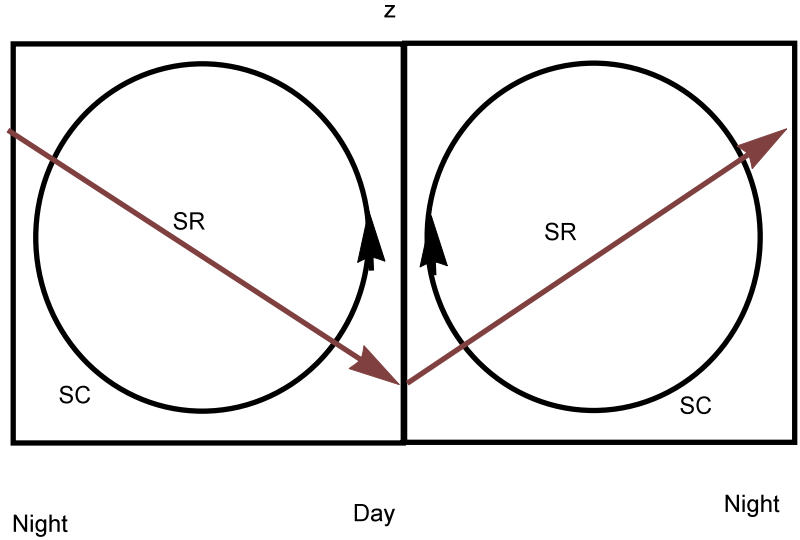}

\caption{Schematic patterns of the bilaterally symmetric-circulation (SC) and
asymmetric super-rotation (SR). The horizontal direction of the plane
represents the equator of the Venus, with right and left ends identified
with each other. The vertical direction represents the height of the
atmosphere. Note the SC respects the bilateral symmetry of the meridian
plane passing through the day point and the two poles, while SR violates
this symmetry. }
\label{fig.1}
\end{figure}

There have been many studies to explain SR from various points of
view. Schubert and Whitehead\cite{Schubert1969}, and Thompson\cite{Thompson1970}
have considered a shift of the day-night convective circulations.
They considered a possible overall shift of the circulations toward
the planet rotational direction that might induce the super-rotation.
The idea would be natural, however, this pair of bilaterally symmetric-circulations
(SC) cannot smoothly connect to SR dominated flow(Fig.\ref{fig.1}).
This is because the flow speeds at a high and low altitude of the
circulations are the same with each other. Therefore SR, that lacks
the circulation at the low altitude, is impossible. 

In order to overcome this difficulty, we introduce an \textit{extra
circulation} rotating around the planet at low altitude. The superposition
of these three circulations can express the SC as well as SR at the
same time. The flow of SC respects the bilateral symmetry about the
meridian section defined by the day point, North/South poles, and
the night point. On the other hand, the flow of SR violates this symmetry.
Therefore our task is to find a mechanism of transition from SC to
SR, \textit{i.e.} the \textit{violation of the symmetry}. 

Subsequently, many researchers considered the \textit{explicit} mechanism
of this symmetry breaking searching for concrete driving forces for
SR. Fels and Lindzen\cite{Fels1974} considered the thermally excited
gravity waves at the cloud top regions. Hou and Farrell\cite{Hou1987}
considered the propagation of the gravity waves upward. Gierasch\cite{Gierasch1975}
considered the systematic shift of the meridian circulations. Matsuda\cite{Matsuda1980,Matsuda1982}
extracts several relevant flow modes and considered their non-linear
interactions based on Gierasch\cite{Gierasch1975}. All of them considered
the \textit{explicit} symmetry breaking based on the principle: Symmetric
mode interactions do not yield asymmetry\cite{Matsuda1983}. Therefore
the planetary rotation was essential for generating SR. 

We recognize, however, the flow of atmosphere is a heat engine system
that autonomously works by getting energy mainly from the Sun and
dumping the heat toward the interstellar space. The basic architecture
of an engine is the linear system composed of a piston and a cylinder.
This linear oscillatory motion of the piston is transformed into the
rotational motion of a wheel either to the right or left. The bilateral
symmetry is spontaneously broken at this stage. Any small trigger
or random fluctuation in the joint or initial condition is needed
to determine the rotational direction. The power of the rotation is
maintained by that of the piston and not by the detailed mechanism
of the joint. Moreover, the time scale for the wheel to reach the
steady state would not depend on the strength of the trigger but depends
on the power generated at the cylinder. 

In this paper, we reconsider the basic mechanism how SR is possible.
We first estimate the number of zonal bands of flow for each planet
in sec.\ref{sec:Number-of-coriolis-drive}. We find the Venus and
the Titan will possibly have a single zone flow. We next introduce
the three-loop model for SC and SR and demonstrate several typical
time evolution of flow patterns in section \ref{sec:Three-loop-model-for}.
In section \ref{sec:Parameter-space-and}, we find stationary points
which correspond to SC and SR with the associated stationary temperature
differences. Next, in section \ref{sec:Physics-behind-and}, we clarify
the Physics behind our model; spontaneous symmetry breaking and the
maximum flow principle. In section \ref{sec:Observational-aspects},
we estimate the parameters in our phenomenological model based on
several observational data. In section \ref{sec:Comparizon-with-other},
we compare our model with other models so far proposed. In the last
section \ref{sec:Conclusions-and-prospects}, we close our study summarizing
our work and prospects.

\section{Number of Coriolis-drive zonal bands\label{sec:Number-of-coriolis-drive}}

The number of atmospheric flow bands rotating around the planet is
determined by the meridional circulation and the Coriolis force. The
Coriolis force makes the meridional circulation velocity $v$ toward
the longitudinal direction of amount $-2\Omega\times v$. Within the
time interval $\tau$, the meridional flow travels the distance $l\approx\tau v$,
and the Coriolis acceleration $\approx2\Omega v$ makes the flow deviate
toward the longitudinal direction about$\approx\left(1/2\right)2\Omega v\tau^{2}$.
If we set this distance as $l$, then we can estimate the distance
for the meridional flow turns its direction about $\pi/2$. Thus we
have $\tau\approx\Omega^{-1}$ and the characteristic distance for
the meridional flow becomes $l\approx v/\Omega$. If we divide the
full meridional distance of the planet by this amount, we can roughly
estimate the number of segments of the meridional circulation, 
\begin{equation}
\#\mathrm{band}=1+\frac{\pi\cdot\left(\mathrm{planetary\,radius}\right)}{v\cdot\left(\mathrm{rotation\,period}\right)}.\label{eq:1}
\end{equation}
This is also the number of (local) super-rotation (SR) bands or jet
streams. It is important to notice that the Coriolis force simply
shifts the existing flow but never accelerates the flow. Reflecting
the fact that the adjacent meridional circulation has the parallel
flow interface (\textit{i.e.} opposite circulation directions), the
zonal bands or jet streams have alternating rotational directions.
The estimated zonal band numbers according to the Eq.(\ref{eq:1})
is given in the Table \ref{tbl1}. 

\begin{table}
$\begin{array}{|c|c|c|c|c|}
\hline \text{Name} & \text{Radius(km)} & \text{Rotation Period(hour)} & \text{Surface Pressure(bars)} & \text{estimated zone number}\\
\hline \text{MERCURY} & 2439.5 & 4222.6 & 0 & 1.00672\\
\hline \text{VENUS} & 6052. & 2802 & 92 & 1.03\\
\hline \text{EARTH} & 6378. & 24 & 1 & 4.09\\
\hline \text{MOON} & 1737.5 & 708.7 & 0 & 1.02853\\
\hline \text{MARS} & 3396. & 24.7 & 0.01 & 2.59977\\
\hline \text{JUPITER} & 71492. & 9.9 & \text{Unknown*} & 85.025\\
\hline \text{SATURN} & 60268. & 10.7 & \text{Unknown*} & 66.5374\\
\hline \text{URANUS} & 25559. & 17.2 & \text{Unknown*} & 18.2903\\
\hline \text{NEPTUNE} & 24764. & 16.1 & \text{Unknown*} & 18.897\\
\hline \text{PLUTO} & 1185. & 153.3 & 0 & 1.08994\\
\hline \text{Titan} & 2575.5 & 382. & 1.45 & 1.07845
\\\hline \end{array}$

\caption{Table of the planet data \cite{NASA2017} and the estimated number
of zonal flow bands from Eq.(\ref{eq:1}). We have taken a representative
value for the flow speed as $v=75$meter/sec. Actual numbers of flow
bands vary in time and may have substructures especially in big planets\cite{Rogers2017}.
The estimated band number qualitatively represents the actual number. }
\label{tbl1}
\end{table}

According to this table, the zonal band number of Mercury, Venus,
Our Moon, Pluto, and Titan are almost one. This fact suggests that
these planets (satellites) have a single zonal band flow or the (global)
SR for each if they have an atmosphere. Consulting the surface pressure,
we expect that Venus and Titan will probably have the (global) SR,
and Mercury, Our Moon, and Pluto are excluded. 

\section{Three-loop model for flows\label{sec:Three-loop-model-for}}

\begin{figure}

\includegraphics{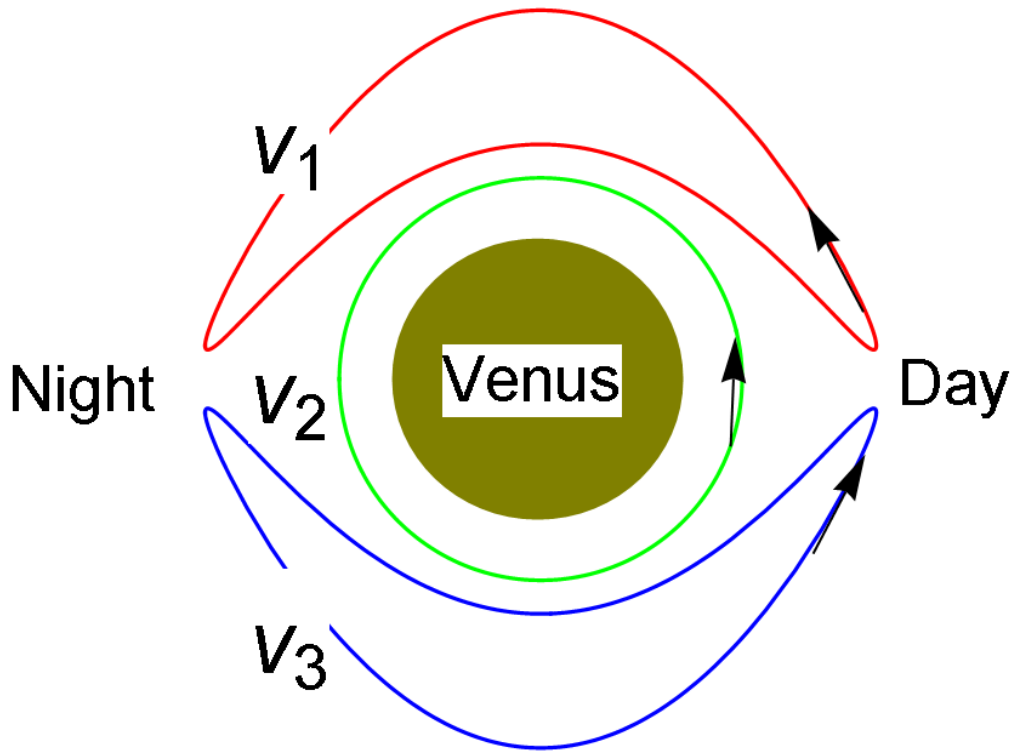}\caption{The three loop model for the Venus atmosphere described by Eq.(\ref{eq:3loop model}).
The arrow defines the positive direction for each circulation. The
superpositions of these flows represent actual flows at high and low
altitude layers. These circulations interact with each other through
viscosity. The loop pair 1-2 and 3-2 tend to be parallel, while the
pair 1-3 tends to be anti-parallel. The total system has two types
of typical flows: +,+,+ (-,-,-), and +,0,- (-,0,+), for the signatures
of $v_{1}\left(t\right),v_{2}\left(t\right),v_{3}\left(t\right)$,
respectively. The former represents SR and the latter SC. }
\label{fig.2}
\end{figure}

We study the planets/satellites with a single zonal band flow such
as Venus and Titan. For this purpose, we introduce the following model
composed of three loops of circulating flows (Fig.\ref{fig.2}). Their
velocities are $v_{1}\left(t\right),v_{2}\left(t\right),v_{3}\left(t\right)$,
and the temperature difference between the day and night points is
$\theta\left(t\right)>0$. These four variables are our basic degrees
of freedom in the model. They obey the set of time evolution equations, 

\begin{align}
\begin{aligned}\frac{dv_{1}}{dt}\text{\ensuremath{=}- \ensuremath{\nu}}v_{1}+R+av_{1}\theta-\kappa\left(v_{1}\text{-}v_{2}\right)\text{-\ensuremath{\frac{\mu}{2}} (}v_{1}\text{+}v_{3}\text{), }\\
\frac{dv_{2}}{dt}=\ensuremath{\kappa}(v_{1}\text{-}v_{2}\text{)+ \ensuremath{\kappa} (}v_{3}\text{-}v_{2}\text{)}\text{, }\\
\frac{dv_{3}}{dt}\text{\ensuremath{=}- \ensuremath{\nu}}v_{3}+R+av_{3}\theta\text{-\ensuremath{\kappa\left(v_{3}\text{-}v_{2}\right)}}-\ensuremath{\frac{\mu}{2}}(v_{3}\text{+}v_{1}\text{), }\\
\frac{d\theta}{dt}\ensuremath{=}h-b(v_{1}^{2}+v_{3}^{2})\theta.
\end{aligned}
\label{eq:3loop model}
\end{align}
Our parameters are the triggering force $R$, temperature drive efficiency
$a$, the energy transfer efficiency $b$, the incoming heat rate
$h$, horizontal viscosity at the lower layer$\kappa$, the vertical
viscosity $\mu$, and the viscosity at the higher layer $\nu$. The
vertical viscosity $\mu$ promotes the bilaterally symmetric-circulations
for the flows $v_{1}$ and $v_{3}$. The horizontal viscosity at the
lower layer$\kappa$ is effective for actual flow $(v_{1}\text{-}v_{2})$
and $\text{(}v_{3}\text{-}v_{2}\text{)}$, and makes the coupling
between all the three flows $v_{1}$ through $v_{3}$. 

All the terms are bilaterally symmetric and do not distinguish east-west
directions except the triggering force $R$. In other words, the set
of equations, except $R$, is invariant under the mirror transformation:
$v_{1}\rightarrow-v_{3},v_{2}\rightarrow-v_{2},v_{3}\rightarrow-v_{1},\theta\rightarrow\theta$.
There are two kinds of typical flows described by the above set of
equations, one respects the symmetry (SC) and the another violates
it (SR). We will see that the latter type of flow SR appears spontaneously
and is drove by the heat $h$ even within the symmetric situation
$R=0$. 

The evolution equations for the flows $v_{i}$ can be written by the
following quasi-potential $V$

\begin{equation}
V=\frac{\kappa}{2}\left(\left(v_{1}-v_{2}\right)^{2}+\left(v_{3}-v_{2}\right)^{2}\right)+\frac{\mu}{4}\left(v_{1}+v_{3}\right)^{2}+\frac{1}{2}\left(\nu-a\theta\right)\left(v_{1}^{2}+v_{3}^{2}\right)-R\left(v_{1}+v_{3}\right)
\end{equation}
as 
\begin{equation}
\frac{dv_{i}}{dt}=-\frac{dV}{dv_{i}},\label{eq:potential derivative}
\end{equation}
for $i=1,2,3$. The first two terms in $V$, each proportional to
the difference/summation of the velocities, represent the momentum
conservation, while the last two terms represent genuine dissipation
and the source term. 

If we neglect the explicit trigger term $R\rightarrow0$, the potential
is entirely quadratic. Therefore, multiplying $v_{i}$ to the both
sides of Eq.(\ref{eq:potential derivative}) and summing over $i=1,2,3$,
we obtain 
\begin{equation}
\frac{dK}{dt}=-\sum_{i}\frac{dV}{dv_{i}}v_{i}=-2V,
\end{equation}
where $K\equiv\left(1/2\right)\sum_{i}\dot{v}_{i}^{2}$ is the strength
of the total flow or the kinetic energy. Thus, the quasi-potential
$V$ drives the total flow. 

We show some results obtained from the numerical calculations of our
model. It will become clear that the typical flow modes, SC or SR,
are mainly determined by the parameters $\kappa$ and $\mu$. Therefore
we first examine the case that the explicit driving is absent $R$=0. 

\begin{figure}
\includegraphics{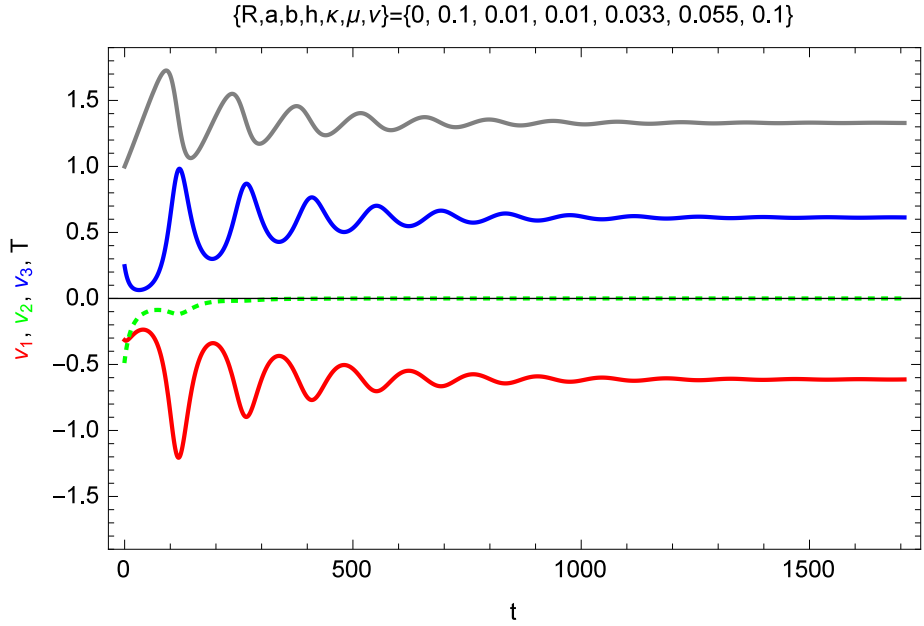}\caption{Time evolution of the three flows $v_{1}$(red), $v_{2}$(green dotted),
$v_{3}$(blue), and the temperature difference $\theta$(gray) in
the SC mode for the case $\kappa<\mu$. The explicit triggering force
is absent $R=0$. }
\label{fig.3}
\end{figure}

\begin{figure}
\includegraphics{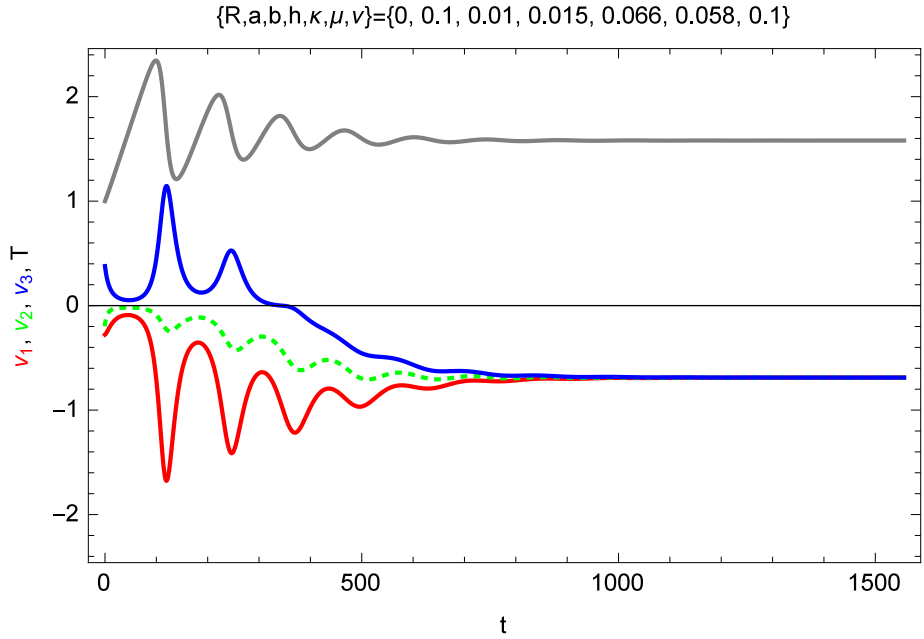}\caption{The same as Fig. \ref{fig.3}, but in the SR mode for the case $\kappa>\mu$.
The explicit triggering force is absent $R=0$. }
\label{fig.4}
\end{figure}

In Fig.\ref{fig.3} and Fig.\ref{fig.4}, we demonstrate that the
parameter $\kappa$ drives SR while $\mu$ drives SC. The SR on Venus
and Titan thus correspond to the case $\kappa>\mu$. 

\begin{figure}
\includegraphics{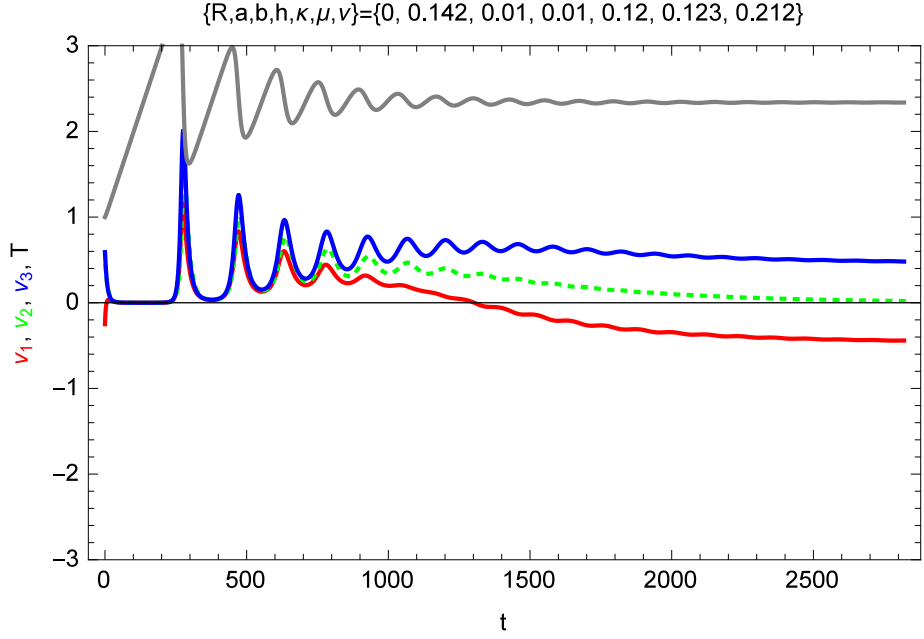}\caption{The same as Fig. \ref{fig.3} but shows the transition from SR to
SC for the case $\kappa\approx\mu$ with $\mu$ slightly larger than
$\kappa$. The explicit triggering force is absent $R=0$.}
\label{fig.5}
\end{figure}

If both the parameters are almost equal $\kappa\approx\mu$, then
the two modes SC and SR compete with each other to yield strong fluctuations
during the transition (Fig.\ref{fig.5}). In this case (Fig.\ref{fig.5}),
SC finally dominates since $\kappa<\mu$. 

\begin{figure}
\includegraphics{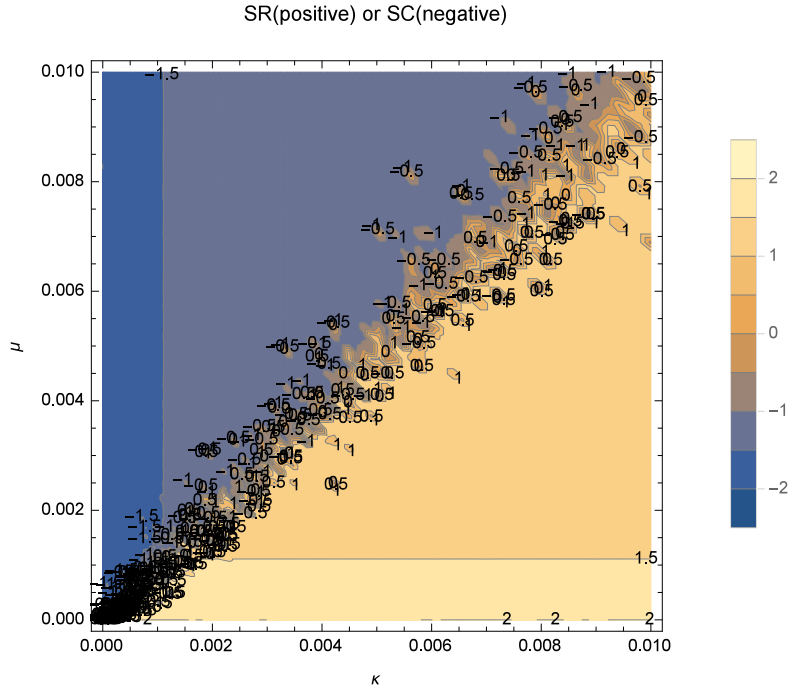}\caption{The final values $v_{1}v_{3}$ at the end of various runs for different
parameters $\kappa$ and $\mu$ with many random initial conditions
($-1<v_{1},v_{2},v_{3}<1$, and $\theta=1$). The other parameters
are fixed to be $R=0,a=0.2,b=0.01,\nu=0.02$, and each run starts
at the time $0$ to $500$. A positive value of $v_{1}v_{3}$ means
the flow 1-2 have the same direction while a negative value of $v_{1}v_{3}$
means they have the opposite direction. The explicit triggering force
is absent $R=0$.}
\label{fig.6}
\end{figure}

In general, the case $\kappa\approx\mu$ is delicate as expressed
in Fig.\ref{fig.6}. The fate of the flow is quite dependent on the
initial conditions and any small fluctuations in the actual case.
In other regions, either SR or SC is realized independently from the
initial conditions. 

\begin{figure}
\includegraphics{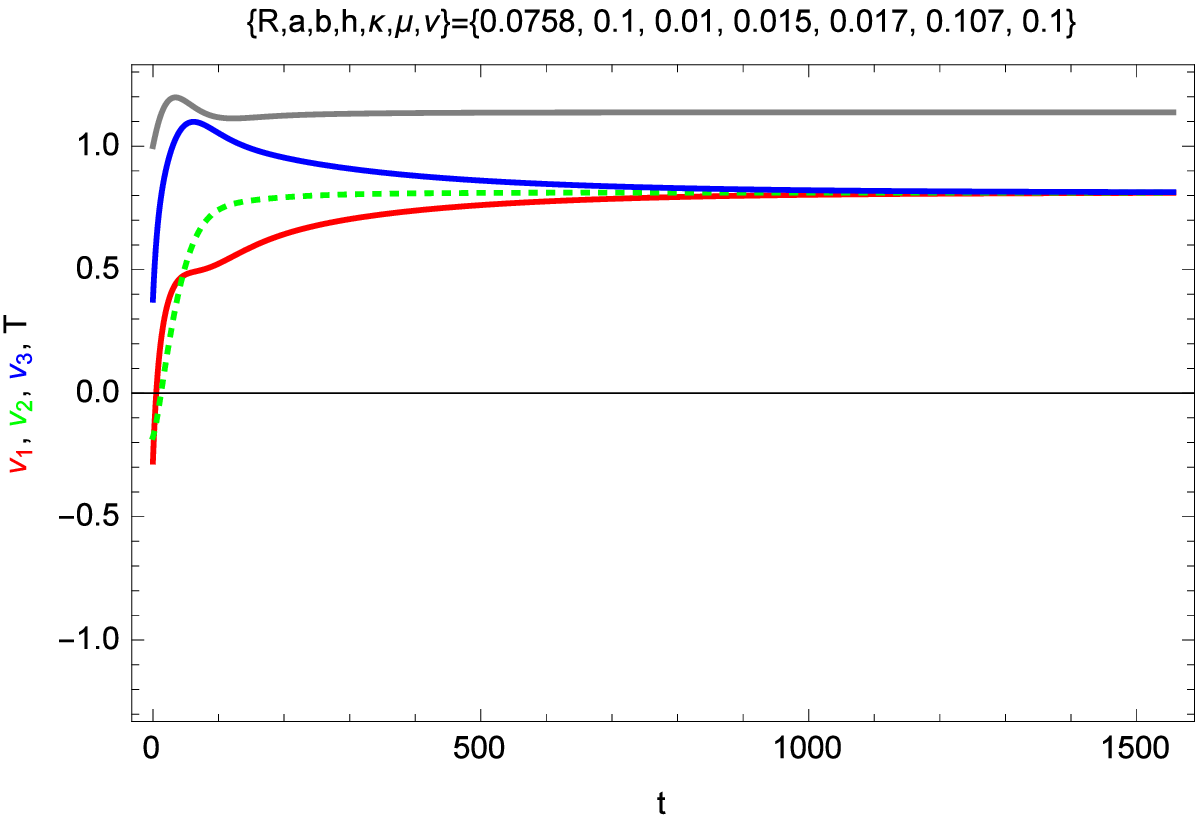}\caption{The same as Fig. \ref{fig.3}, but now we have the explicit triggering
force $R$ which drives SR. In this case, the effect of the force
$R$ dominates even if $\kappa<\mu$.}
\label{fig.7}

\end{figure}

If the explicit triggering force $R$ exists, the situation is quite
different (Fig.\ref{fig.7}). SR flow is directly triggered almost
independently from the values of $\mu$ and $\nu$. The the local
SR on the Earth and Jupiter will correspond to this case $R>0$. 

\section{Parameter space and the stationary states\label{sec:Parameter-space-and}}

We clarify the characteristic behaviors of the flow evolution demonstrated
in the previous section. The relevant parameters of our model Eq.(\ref{eq:3loop model})
are...$\kappa,\mu,\nu$ for the interactions between flows, and $a,b$
for flow-heat coupling, and $h,R$ for the driving force. We will
find the possible fixed stationary points of Eq.(\ref{eq:3loop model})
for the variables $v_{1},\;v_{2},\;v_{3},\;\theta$. Setting the left-hand
sides $0$ in this set of equations and solving for these variables,
the general fixed points are found to be, in the form of $\{v_{1},\;v_{2},\;v_{3},\;\theta\}$,
\begin{equation}
\begin{array}{cccc}
\{v_{-}^{SR}, & v_{-}^{SR}, & v_{-}^{SR}, & \theta_{+}^{SR}\},\\
\{v_{+}^{SR}, & v_{+}^{SR}, & v_{+}^{SR}, & \theta_{-}^{SR}\},\\
\{v_{-}^{SC}, & -\frac{R}{\kappa-\mu}, & v_{+}^{SC}, & \frac{\kappa+\nu}{a}\},\\
\{v_{+}^{SC}, & -\frac{R}{\kappa-\mu}, & v_{-}^{SC}, & \frac{\kappa+\nu}{a}\},
\end{array}\label{eq:stationary-sol1}
\end{equation}
where
\begin{align}
v_{\pm}^{SR} & \equiv\frac{R\pm\sqrt{2\left(a/b\right)h\left(\mu+\nu\right)+R^{2}}}{2\left(\mu+\nu\right)},\;v_{\pm}^{SC}\equiv\frac{\pm\frac{\sqrt{\left(a/b\right)h\left(\kappa-\mu\right)^{2}-2R^{2}(\kappa+\nu)}}{\sqrt{2}\sqrt{\kappa+\nu}}-R}{\left(\kappa-\mu\right)},\\
\theta_{\pm}^{SR} & \equiv\frac{\pm\sqrt{b}R\sqrt{2ah\left(\mu+\nu\right)+bR^{2}}+ah\left(\mu+\nu\right)+bR^{2}}{a^{2}h}.
\end{align}
These four fixed points are too complicated. Therefore we reduce the
expression assuming the relevant case $R\rightarrow0$ where the evolution
equation Eq.(\ref{eq:3loop model}) is fully symmetric. Then Eq.(\ref{eq:stationary-sol1})
reduces to the form 
\begin{equation}
\begin{array}{cccc}
\{-\sqrt{\frac{ah}{2b\left(\mu+\nu\right)}}, & -\sqrt{\frac{ah}{2b\left(\mu+\nu\right)}}, & -\sqrt{\frac{ah}{2b\left(\mu+\nu\right)}}, & \frac{\mu+\nu}{a}\},\\
\{\sqrt{\frac{ah}{2b\left(\mu+\nu\right)}}, & \sqrt{\frac{ah}{2b\left(\mu+\nu\right)}}, & \sqrt{\frac{ah}{2b\left(\mu+\nu\right)}}, & \frac{\mu+\nu}{a}\},\\
\{-\sqrt{\frac{ah}{2b\left(\nu+\kappa\right)}}, & 0, & \sqrt{\frac{ah}{2b\left(\nu+\kappa\right)}}, & \frac{\kappa+\nu}{a}\},\\
\{\sqrt{\frac{ah}{2b\left(\nu+\kappa\right)}}, & 0, & -\sqrt{\frac{ah}{2b\left(\nu+\kappa\right)}}, & \frac{\kappa+\nu}{a}\}.
\end{array}\label{eq:stationary-solR=00003D0}
\end{equation}
These four fixed points represent, in order, left and right rotating
SR, and left and right circulating SC (Fig.\ref{fig.2}). The choice
of either the left or the right rotating SR violates the bilateral
symmetry of the Eq.(\ref{eq:3loop model}), while SC respects the
symmetry. We find again that the distinction between SR and SC is
summarized in the parameters $\mu,\nu$ as apparent in the above expressions.
We also find the flow speed squared is always inversely proportional
to the temperature $v^{2}=\left(h/2b\right)/\theta$. We further find
that the triggering force term $R$ turns out to kill the SC mode
for $R>R_{c}$ where
\begin{equation}
R_{c}=\left(\frac{ah\left(\kappa-\mu\right)^{2}}{2b\left(\kappa+\nu\right)}\right)^{1/2}
\end{equation}
since $v_{\pm}^{SC}$ becomes complex then. 

If the entire flow were due to the triggering force $R$, then the
symmetry is explicitly broken. Putting $h\rightarrow0$, we have the
stationary points, for $\{v_{1},\;v_{2},\;v_{3},\;\theta\}$, as 
\begin{equation}
\begin{array}{cccc}
\{0, & 0, & 0, & *\},\\
\{\frac{R}{\left(\mu+\nu\right)}, & \frac{R}{\left(\mu+\nu\right)}, & \frac{R}{\left(\mu+\nu\right)}, & *\},\\
\{v_{-}^{SC}, & \frac{R}{\mu-\kappa}, & v_{+}^{SC}, & *\},\\
\{v_{+}^{SC}, & \frac{R}{\mu-\kappa}, & v_{-}^{SC}, & *\},
\end{array}\label{eq:stationary-sol h=00003D0}
\end{equation}
where the velocity of the symmetric flow SC always becomes complex
\begin{equation}
v_{\pm}^{SC}\equiv\frac{\left(1\pm i\right)}{\kappa-\mu}R
\end{equation}
and $"*"$ means being indeterminate. Thus the mode SC is fully killed
and only the SR survives whose speed is simply proportional to the
force $R$. 

We now study the stability of the fixed points obtained above. The
right-hand side of Eq.(\ref{eq:3loop model}) linearized, in the coordinate
$\left(\begin{array}{cccc}
v_{1} & v_{2} & v_{3} & \theta\end{array}\right)$, becomes

\begin{equation}
\left(\begin{array}{cccc}
\begin{array}{c}
a\theta-\kappa-\frac{\mu}{2}-\nu\\
\kappa\\
-\frac{\mu}{2}\\
av_{1}
\end{array} & \begin{array}{c}
\kappa\\
-2\kappa\\
\kappa\\
0
\end{array} & \begin{array}{c}
-\frac{\mu}{2}\\
\kappa\\
a\theta-\kappa-\frac{\mu}{2}-\nu\\
av_{3}
\end{array} & \begin{array}{c}
-2b\theta v_{1}\\
0\\
-2b\theta v_{3}\\
-b\left(v_{1}^{2}+v_{3}^{2}\right)
\end{array}\end{array}\right).\label{eq:linear dev}
\end{equation}
The eigenvalues of this matrix evaluated by the solutions Eqs.(\ref{eq:stationary-sol1},\ref{eq:stationary-solR=00003D0})
give the instability of the individual solutions. If we set $R=0$
and using Eq.(\ref{eq:stationary-solR=00003D0}), we get the following
instability by numerical calculations. For the case $\mu>\kappa$,
the first two lines of solutions of Eq.(\ref{eq:stationary-solR=00003D0})
(SR) give 1 positive and 3 negative eigenvalues, with some imaginary
part in the latter. This means SR mode is unstable with damping oscillatory
behavior. On the other hand the last two lines of solutions of Eq.(\ref{eq:stationary-solR=00003D0})
(SC) give 4 negative eigenvalues, with some imaginary part in there.
This means SC mode is stable. In this case, the velocity of SC mode
is faster than SR mode and the temperature difference $\theta$ for
SC is smaller than that for SR as is seen in Eq.(\ref{eq:stationary-solR=00003D0}). 

For the opposite case $\mu<\kappa$, SR gives 4 negative eigenvalues
while SC gives 1 positive and 3 negative eigenvalues. This means that
SR mode is stable. In this case, SR mode is faster and $\theta$ is
smaller than those of SC mode. 

Summarizing the (in)stability, we conclude that \textit{the flow always
chooses the faster mode, or the flow always choose the lower temperature}.
This also means that \textit{the flow automatically chooses the most
efficient mode of heat transfer}. 

\section{Physics behind the generation of SR mode\label{sec:Physics-behind-and}}

We have introduced a simplest model Eq.(\ref{eq:3loop model}) for
describing the SR excluding any secondary effects as much as possible.
This is because we want to clarify the physics of SR before the elaborate
study on the individual detail. Therefore in this section, we study
basic physics behind our model. 

We first emphasize that the model Eq.(\ref{eq:3loop model}) is minimal
and simplest in the following sense. 
\begin{enumerate}
\item (three loops) The model is composed of three flow loops of velocities
$v_{1},v_{2},v_{3}$. If it were two-loop, the two kinds of flow modes
SR and NR cannot be described in the same model. This is because the
speeds of the upper and lower region have the same amount all the
time\cite{Thompson1970}. This is not the property of SR. 
\item (dynamical temperature) The temperature difference $\theta(t)$ appears
as a dynamical variable. If it were a constant parameter, then the
model does not show the spontaneous symmetry breaking. Even the multiply
coupled Lorenz model \cite{Lorenz1963} is not enough for our purpose
because the external temperature difference is fixed from the beginning.
The flow that carries heat must have feedback effect for the reduction
of the source temperature. 
\item (non-linearity) Our model has nonlinear terms that come from the velocity-temperature
couplings $av_{1}\theta,av_{3}\theta,$and $b(v_{1}^{2}+v_{3}^{2})\theta$.
These couplings have been chosen so that it does not destroy the bilateral
symmetry of the system. The ultimate origin of them would be the advection
term. However, there will appear many nonlinear terms in actual situations,
such as the velocity dependent convection and turbulent fluctuations.
Provided they respect the symmetry, these nonlinear terms will not
drastically change our model as we have already checked some of them.
We believe our nonlinear terms are minimal. 
\item (extra loop under the symmetric loops). We set an extra rotating loop
$v_{2}$ \textit{under} the symmetric-circulation loops $v_{1},v_{3}$.
If this extra loop were set \textit{over} the symmetric loops, then
$v_{1},v_{3}$ eventually reduce for the horizontal friction near
the planet surface. This further reduces $v_{2}$ as well for the
loop couplings. Setting the extra loop under $v_{1},v_{3}$ helps
the reduction of the overall speed under circulations, $v_{1}-v_{2},v_{3}-v_{2}$,
and leaving high-speed outer rotation $v_{1},v_{3}$. 
\end{enumerate}
There are two basic physics in our model; spontaneous symmetry breaking
(SSB) and the possible maximum flow principle (MFP). 

\begin{figure}
\includegraphics{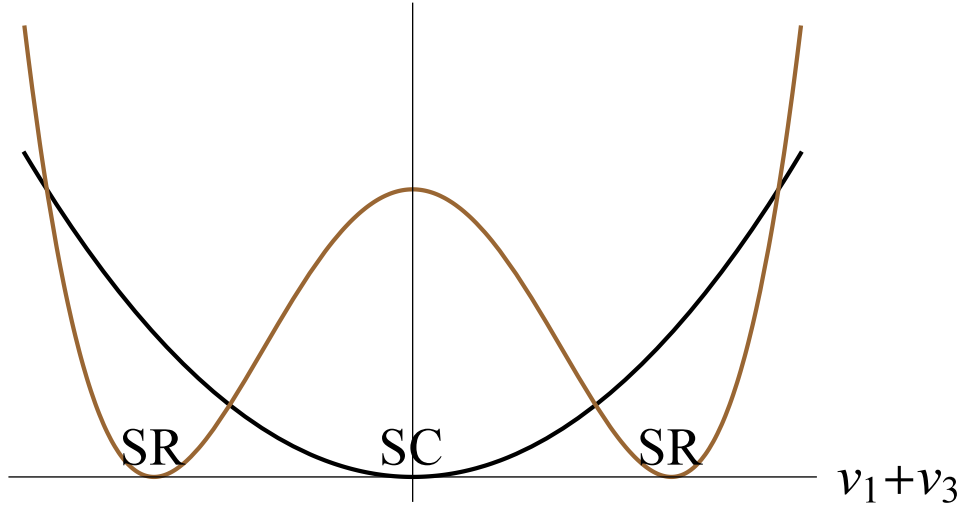}\caption{Schematic diagram of the spontaneous breaking of the bilateral symmetry.
The horizontal axis represents any order parameter, the indicator
of the bilateral symmetry violation, say the combination of the variables
$v_{1}+v_{3}$. The vertical axis represents any indicator of the
stability, the lower the position the more stable, say the temperature
difference $\theta$. The flow mode SC respects the symmetry while
SR doesn't. The flow mode SR appears spontaneously without resorting
to the explicit symmetry violating force such as $R$.}
\label{fig.8}
\end{figure}

Spontaneous symmetry breaking (SSB) is very general phenomena we encounter
everywhere. Our model Eq(\ref{eq:3loop model}) respects the bilateral
symmetry and does not distinguish east-west directions ($v_{1}\rightarrow-v_{3},v_{2}\rightarrow-v_{2},v_{3}\rightarrow-v_{1},\theta\rightarrow\theta$),
except the triggering force $R$ which we neglect for the moment.
However, the symmetry can be violated at the solution level. This
situation is schematically shown in Fig.\ref{fig.8}. The horizontal
axis represents any order parameter, the indicator of the bilateral
symmetry violation, say $v_{1}+v_{3}$. The vertical axis represents
any indicator of the stability, the lower the position the more stable,
say the temperature difference $\theta$. The flow SC respects the
symmetry while SR violates this symmetry. This violation mode SR can
appear spontaneously whenever this mode is more stable than the others. 

The planetary atmosphere is a heat engine. The heat injection from
the sun yields work from the system and the remaining heat flows out
into space. Any tiny trigger or any random initial condition, as well
as the explicit trigger $R$, can decide the initial rotational direction
whichever. The non-linearity of the system enhances the rotation to
this direction. This mechanism is the same as the ordinary engine
with a linear cylinder and piston. The linear periodic motion of the
piston and the rod gradually enhance and establish the rotational
motion of either direction. 

The maximum flow principle (MFP) is another essence of our model.
There are multiple modes for the heat propagation in general heat
system. Depending on the boundary or initial conditions, the system
often chooses the most efficient mode for heat transfer autonomously.
In our case, we have typically SC and SR modes. SC respects the bilateral
symmetry and SR violates it. As we have seen at the end of the last
section, when the parameters satisfy $\kappa<\mu$, SC mode is autonomously
chosen. This SC flows faster and therefore the temperature difference
reduces, thus is more efficient than SR mode. On the other hand, when
$\kappa>\mu$, more efficient flow SR is chosen. 

This tendency of MFP seems to be common in various cases. For example
for water boiling, the heat transfer modes changes, in the order of
increasing temperature difference, conduction, convection, nucleate
boiling, and passing through the phase of transition boiling, finally
reaches to the film boiling\cite{Nukiyama1934}. The efficiency of
the heat transfer actually increases in this order. MFP will be one
of the common rules that govern the non-equilibrium phenomena of fluid.
However, the theoretical formulation of this kind of variational principle
for thermo-fluid dynamics far from equilibrium seems not to be established
yet\cite{Onsager1953}. 

Finally, I should mention that the above SSB and MFP are closely related
to each other. Ordinary SSB requires an effective potential that determines
the stability, while in our case, we do not have such potential. The
best indicator of stability would be the flow efficiency or the temperature
$\theta$. In this sense, our SSB is special and is supported by the
MFP. 

\section{Observational aspects\label{sec:Observational-aspects}}

We now try to constrain our model parameters from available observational
data. The vertical temperature profile of Venus atmosphere is observed
\cite{Seiff1983} and the fluctuations are not small especially in
the higher layers: at the height $65km$, the temperature is $220-250K$
and at $90km$ and higher, $160-190K$. If we can interpret the fluctuation
as the day and night temperature difference $\theta$ at the SR zone
in the mesosphere, it becomes $\theta_{M}=30K$ at most. Another possibility
would be that the day and night temperature difference $\theta$ comes
from the difference in heights for the SR. If the SR layer were higher
at the night region (colder) and were lower at the day region (hotter),
then we have $\theta=60K$ at most. Suppose we take $\theta_{M}=30K$
below. On the other hand in the much higher thermosphere, higher than
120km, $\theta_{T}=150K$ where the flow may be SC\cite{Bertaux2007}.
Therefore we have 
\begin{equation}
\frac{\theta_{T}}{\theta_{M}}\approx\frac{150K}{30K}=5.
\end{equation}
According to Eq.(\ref{eq:stationary-solR=00003D0}) at the stationary
points, we have the relationship between the wind speed and the temperature,
\begin{equation}
v^{2}=\frac{ah}{2b\theta}
\end{equation}
irrespective of SR or SC. Then from the temperature ratio, we can
estimate the wind speed ratio, assuming $a,b$ are respectively the
same at thermosphere and mesosphere, 
\begin{equation}
\left(\frac{v_{T}}{v_{M}}\right)^{2}=\frac{\theta_{M}}{\theta_{T}}\frac{h_{T}}{h_{M}}=\frac{1}{5}\frac{h_{T}}{h_{M}}.
\end{equation}
Since we naturally expect, from the existence of the heavy cloud of
$H_{2}SO_{4}$, that $h_{M}>h_{T}$, we estimate $v_{T}$ is much
smaller than $v_{M}$. 

Observation shows that the flow is SR in the mesosphere, and therefore
we expect $\kappa>\mu$ there. On the other hand at the higher thermosphere,
the horizontal viscosity $\kappa$ would be far smaller. Therefore
the opposite situation $\kappa<\mu$ would be probable there. Then
the flow would be SC. At the interface zone, $\kappa\approx\mu$ and
the frustration between SR and SC would take place. Therefore flows
there would have strong fluctuation as in Fig.\ref{fig.6}. 

We now estimate the parameters of SR zone in the mesosphere of Venus.
The main ingredient there is the sulfuric acid $H_{2}SO_{4}$ which
has the specific heat $c=1.4$(J/g K). The mass density of the sulfuric
acid in the mesosphere is $\rho=4.4\times10^{2}g/m^{3}$, and the
solar energy input is $W=150Watt/m^{2}$. The specific height of this
zone is $H=3.0\times10^{4}m$. Thus we have 
\begin{equation}
h=\frac{d\theta}{dE}\frac{dE}{dt}=\text{\ensuremath{\frac{W}{c\rho H}=8.1\times10^{-6}\frac{K}{\sec}}}.
\end{equation}
Using this value and observed wind velocity $v=100m/\sec$, we have
\begin{equation}
b=\frac{h}{2\theta v^{2}}=1.3\times10^{-11}\frac{\sec}{m^{2}}.
\end{equation}
The time scale of the temperature difference $\theta$ therefore becomes
\begin{equation}
\tau_{\theta}=\frac{1}{bv^{2}}=\frac{2\theta}{h}=7.4\times10^{6}\sec.
\end{equation}
The similar time scale for the wind velocity can be roughly estimated
to be 

\begin{equation}
\tau_{\nu}=\frac{1}{\nu}=4.7\times10^{8}\sec
\end{equation}
from the observations \cite{Khatuntsev2013,Kouyama2013} claiming
that the SR velocity had changed $40m/\sec$ within 6 years. If we
suppose the parameters $\mu$ and $\nu$ have almost the same order,
\begin{equation}
a=\frac{\mu+\nu}{\theta}\approx\frac{\nu}{\theta}=7.0\times10^{-11}K^{-1}\sec^{-1}.
\end{equation}

On the other hand, the Saturn's satellite Titan is a candidate of
the global SR according to the argument of the section \ref{sec:Number-of-coriolis-drive}.
However, the study for the Titan SR seems to be complicated by relatively
large deviation of the spin axes $26.7^{o}$. Therefore we skip the
case of Titan SR in this paper. 

\section{Comparison with other models\label{sec:Comparizon-with-other}}

We compare our model with the others so far proposed to explain SR
mainly of Venus. 

We have set up three loop model to describe SR and SC. On the other
hand, Schubert and Whitehead\cite{Schubert1969}, and Thompson\cite{Thompson1970}
proposed a model of two loops, representing the day-night convective
circulations, that are shifted by the planetary rotation. The time
lag, in their case, due to the propagation of the temperature difference
upward, is expected to shift the day-night convective circulation.
Thus the planetary rotation is essential to maintain SR. However,
a simple shift of SC does not fully describe SR whose flow covers
the entire planet surface. We believe a set of three loops is indispensable
to describe SR as well as SC. 

We have emphasized the spontaneous generation of SR in which we do
not need an explicit violation of bilateral symmetry of west-east
mirror reflection. Therefore the planetary rotation is not essential
to yield and maintain SR in our model. On the other hand, most of
the other models explicitly breaks the symmetry for example by the
thermally excited gravity waves at the cloud top regions (Fels and
Lindzen\cite{Fels1974}), or by the propagation of the gravity waves
upward (Hou and Farrell\cite{Hou1987}), or by the systematic shift
of the meridian circulations (Gierasch\cite{Gierasch1975} and extended
by Matsuda\cite{Matsuda1980,Matsuda1982}). All of them considered
the \textit{explicit} symmetry breaking based on the principle: Symmetric
mode interactions do not yield asymmetry\cite{Matsuda1983}. Therefore
the planetary rotation is ultimately essential for generating SR in
their cases, quite contrary to our model. However, the planetary rotation
or the Coriolis force may trigger the symmetry breaking especially
for the cases of Venus and Titan. 

We have extracted the essential circulation modes and try to analyze
them in the analytic method. This point is parallel to Matsuda \cite{Matsuda1980},
Yamamoto\cite{Yamamoto2013}, Kashimura and Yoden \cite{Kashimura2015}.
They developed simplified model extending the mechanism of Gielash\cite{Gierasch1975}.
Their relevant parameters roughly correspond to ours. For example,
the thermal Rossby number $R_{T}=gH\Delta_{H}a^{-2}\Omega^{-2}$ would
correspond to our parameter $h$, the vertical Ekman number $E_{V}=\nu_{V}a^{-2}\Omega^{-1}$to
$\mu$, the horizontal Ekman number $E_{H}=\nu_{H}a^{-2}\Omega^{-1}$
to $\nu$, where $H$ is the height of the top boundary, $\Delta_{H}$
is the fractional change in potential temperature from the equator
to the pole, $a$ is the planetary radius, $\Omega$ is the angular
velocity of the planetary rotation, $g$ is the gravitational acceleration,
$\nu_{H}$ and $\nu_{V}$ are respectively the horizontal and vertical
diffusion coefficients. However, the correspondence is not complete
for example, we do not normalize the parameters by $\Omega$. 

We have derived the expression for SR for Venus as Eq.\ref{eq:stationary-solR=00003D0},
and the SR velocity is given by 
\begin{equation}
v_{SR}=\sqrt{\frac{ah}{2b\left(\mu+\nu\right)}},
\end{equation}
which is independent of the explicit driving force $R$ or $\Omega$;
SR appears spontaneously. Therefore SR would be observed even for
the static planet without rotation $\Omega=0$, or more precisely
the planet with fixed day-night hemispheres. 

On the other hand, Gierasch\cite{Gierasch1975} obtained the SR velocity
as 
\begin{equation}
U=\Omega a\exp\left(D^{2}HW/\nu_{v}\right),
\end{equation}
where $\Omega$ is the angular velocity, $a$ is the Venus radius,
$D$ is the mean scale height, $H$ is depth in scale heights, $W$
is the inverse of the meridional overturning time, and $\nu_{v}$
is the vertical eddy diffusivity. It is apparently proportional to
the angular velocity and $U$ vanishes for no planetary rotation;
SR is generated by the explicit symmetry breaking force. However,
the force $R$ would trigger the SR of Venus and Titan, and becomes
essential to drive the local SR for the other planets such as Earth
and Jupiter. 

\section{Conclusions and prospects\label{sec:Conclusions-and-prospects}}

We constructed a three loop model which describes super-rotation (SR)
as well as symmetric-circulation (SC) of the planetary atmosphere.
The set of equations describing the time evolution of the circulation
velocities $v_{1}\left(t\right),v_{2}\left(t\right),v_{3}\left(t\right)$
and the temperature difference $\theta\left(t\right)$ respects the
bilateral mirror symmetry about the meridian section defined by the
day/night points and poles, except terms of $R$. We demonstrated
that the asymmetric flow pattern SR is spontaneously generated in
our model as well as the symmetric SC. Our model is minimal which
has this property. These SR and SC modes are generally in frustration
with each other and the faster flow, or much efficient flow, is spontaneously
realized depending on the given parameters. In general, the planetary
atmosphere is a thermal engine and the zonal rotation flow is naturally
generated irrespective of globally or locally. We could constrain
the parameters of our phenomenological model. Many sophisticated mechanisms
so far studied, such as the mechanism based on the night-day circulation,
on the meridian circulation, on the thermally excited gravity waves,
will be important to trigger SR and to determine the direction of
SR. However, the SR is quite general and can be spontaneously generated
irrespective of the detail of trigger. 

We did not consider the intrinsic fluctuations of the flow and therefore
our model is deterministic. Actually, the atmospheric circulation
of a planet is a huge system including many degrees of freedom. Therefore
the random fluctuations must exist on top of the dynamics of relevant
variables. This fluctuation effect would be easily realized by introducing
some appropriate random force term in our model. This may cause an
intermittent transition between SR and SC modes in some situations.
Large fluctuations would be crucial in particular for describing the
local SR of the Earth and the Jupiter. 

We did not consider the complete physical derivation of our model
but it was simply proposed phenomenologically. A natural method would
be to extend the Lorenz model to include higher harmonic modes. It
may also work if we couple three Lorenz models. In either case, we
need to include full feedback to the temperature which was absent
in the original Lorenz model, in which only small fluctuations allowed
around a fixed linear temperature gradient. Furthermore, we need to
describe the origin of the non-linearity of the flow which was essential
for describing the spontaneous SR generation in the present paper.

We would like to report soon our further study on SR developing the
improved model reflecting the above points. 
\begin{acknowledgments}
The author would like to thank Hideaki Mouri (Meteorological Research
Institute) for fruitful discussions and Takayoshi Ootsuka (Ochanomizu
University) for careful examinations of the set of equations. The
author also would like to thank all the members of the Ochanomizu
Astrophysics laboratory for their encouragements on this non-standard
discussions. 
\end{acknowledgments}

\end{document}